\documentclass[11pt]{article}
\usepackage[utf8]{inputenc}
\usepackage{algorithm}
\usepackage{algpseudocode}
\usepackage{subcaption}
\usepackage{graphicx}
\usepackage{enumerate}
\usepackage{multirow}
\usepackage{hyperref}
\usepackage{authblk}

\title{Privacy-Aware Crowd Labelling for Machine Learning Tasks}

\author{Giannis Haralabopoulos\thanks{Henley Business School,
        University of Reading,
        i.haralabopoulos@henley.ac.uk}        
\and    
    Ioannis Anagnostopoulos\thanks{Department of Computer Science and Biomedical Informatics,
        University of Thessaly,
        janag@dib.uth.gr}
        }%

\begin{document}

\maketitle

\begin{abstract}

The extensive use of online social media has highlighted the importance of privacy in the digital space. As more scientists analyse the data created in these platforms, privacy concerns have extended to data usage within the academia. Although text analysis is a well documented topic in academic literature with a multitude of applications, ensuring privacy of user-generated content has been overlooked. Most sentiment analysis methods require emotion labels, which can be obtained through crowdsourcing, where non-expert individuals contribute to scientific tasks. The text itself has to be exposed to third parties in order to be labelled. In an effort to reduce the exposure of online users' information, we propose a privacy preserving text labelling method for varying applications, based in crowdsourcing. We transform text with different levels of privacy, and analyse the effectiveness of the transformation with regards to label correlation and consistency. Our results suggest that privacy can be implemented in labelling, retaining the annotational diversity and subjectivity of traditional labelling.

\end{abstract}


\section{Introduction}

Sentiment analysis is a human centered task where emotions are uncovered from information. Modern methods can work with almost any type of emotion-evoking information like multimedia content or images \cite{krishna2017visual,laghari2017systematic}. Most modern emotion models rely on simple textual information found in OSNs (Online Social Networks) or online review sources \cite{kross2018does,xue2018deep,imran2018processing,cambria2017sentiment,wagh2018twitter}.

Text collections are labeled and analysed to create emotion detection and prediction algorithms \cite{haralabopoulos2020ensemble}. Labelling can happen at paragraph, sentence, or term group level. Emotion detection and prediction in lexicon based analysis, start at word level. Words are matched to an emotion using a predefined lexicon (sentiment lexicon) often created through crowdsourcing \cite{kiritchenko2017capturing,deng2018topic,strapparava2004wordnet}. Crowdsourcing enables researchers to reach a wide range of non expert individual contributors \cite{tucci2018creating,basson2018crowdsourcing}, using various platforms \cite{peer2017beyond}, for any task that requires human computation. In internal crowdsourcing, the scientist themselves annotate the corpus, and in external crowdsourcing non expert contributors perform the annotation.

A group of words may comprise a full sentence or a part of a sentence. The sentiment of each word group can be defined without knowledge of the individual word sentiment. Supervised lexicon methods use individual word sentiment to perform initial sentiment classification in texts, and sentiment labels from groups to improve their performance \cite{maas2011learning,pang2008opinion}. The quality of a learning method depends on multiple factors such as the lexicon, the number of labels, and model. When dealing with OSNs, the group of words to label is an OSN submission \cite{medhat2014framework} as a whole. A commonly used practice that overlooks user privacy is when the group of words is provided unaltered to contributors, which makes the creator of the submission potentially traceable.

Data submitted in Social Networks is owned by the Social Network itself and is open data for any interested individual. This should not reduce our share of responsibility to ethically handle that data. We consider online data as personal data and therefore our study aims to further reduce data exposure. We propose a method for masking text elements based on specific text properties, used as transformation agents. This introduces a layer of privacy between the social media users, whose submissions are used in a crowdsourcing task, and the crowd contributors that annotate these submissions. We assess the feasibility of individual terms, but the same method can be applied in a group of terms or whole sentences. Although lexicon based methods and individual term labelling are governed by a certain level of decontexualisation and their meaning might be misinterpreted \cite{puschmann2018turning}, they are the simplest ingredient of supervised sentiment analysis.

As mentioned, the transformations we propose are based on textual properties. In sentiment analysis, these properties are the emotions conveyed through text. In marketing, for example, text can be masked from service providers or third parties to preserve users' privacy. We demonstrate the effectiveness of masking in emotion labeling, where each term corresponds to a range of emotions, and experiment with four different text transformations of varying levels of privacy. We explore the results of these transformations with regards to the emotional diversity contributed and suggest an aggregation of individual term emotions as a validator for sentence labels. The results are compared to usual text annotation, to outline the similarities or differences of subjective privacy-aware labelling versus traditional labelling.

Given a text corpus retrieved from social media, how can we crowdsource labels without exposing social media submissions to crowd workers? Furthermore, how can we evaluate the privately obtained labels with no knowledge of the classic text label? Our contributions to the above problems are: the introduction of a privacy layer between information and crowd workers that retains text properties and the demonstration of per term aggregation as a sentiment evaluation method. Since our text transformation method is quite novel and no similar study exists, we will focus on the most significant crowdsourcing studies that address any form of privacy concern.

\section{Related Work}

Although the privacy of OSNs has been extensively studied since the early 2000s, the privacy of OSN users with regards to the analysis of their social media submissions is relatively unexplored. Researchers have assessed user vulnerabilities in social media and their actions and those of their social circle \cite{gundecha2012mining,gross2005information,zhou2008preserving}, noting that users' privacy may be easily infringed \cite{mitrou2014social} even when scientists and developers use data for fair purposes, such as creating personalized experiences. In \cite{zheng2018fair}, the authors propose the segregation of privacy concerns to sets of varying privacy priority. 

The privacy paradox, as introduced by Banrnes \cite{barnes2006privacy}, and its applicability to OSN users is the subject of \cite{dienlin2015privacy}. The study highlights the correlation of online privacy attitudes with personal privacy attitudes, and concludes that online privacy should no longer be considered as paradoxical. In \cite{song2018personal}, authors propose a taxonomy-guided model to predict non-privacy aware parts of a post and construct guidelines on social media privacy. As privacy in online social networks has always been a concern, it is important to define what constitutes failure to preserve privacy, aptly noted in \cite{amintoosi2015trust}.

Smart living has brought interconnected devices to our daily lives, along with the need for privacy in the IoT space. At a hardware level, \cite{de2011short} introduces a privacy-enhanced participatory query infrastructure for devices and users. Authors of \cite{song2016privacy} present a privacy preserving communication protocol for IoT applications with: Confidentiality, Data Integrity and Authentication features, and \cite{sharma2017cooperative} uses crowdsourcing to preserve privacy rules in the social-IoT space.

Crowdsourcing, the process of obtaining information from non-expert individuals (participators, workers or contributors), can be used towards preserving the privacy within applications, but the privacy of the crowdsourcing task has to be considered as well \cite{diamantopoulou2018assessment}. Privacy-enhancing technologies for analysing personal data are also proposed in \cite{yang2015security}, which focus on Mobile Crowdsourcing Networks. Spatial crowdsourcing is studied in \cite{li2017spatial} where authors employ an encryption of coordinates to preserve location privacy in geometry based tasks. A more indirect use of crowdsourcing towards privacy enhancement is detailed in \cite{zhang2017privacy}, where crowdworkers train a privacy preserving deep computation model for IoT data.

Contributors are also engaged to assess privacy, especially in computer vision: in \cite{korshunov2012crowdsourcing} crowdworkers compare blurring, pixelisation, and masking video effects with regards to privacy, and privacy intrusiveness of HDR imaging is studied in \cite{korshunov2014crowdsourcing} with crowdsourced evaluation. Authors of \cite{wu2018towards} experiment on privacy-preserving action recognition. Crowd sourced OSN data published "as is" poses privacy threats for the participating individuals. The authors of \cite{wang2016real} propose a privacy preserving framework for real-time crowd sourced spatio-temporal data. Databox explores privacy-aware digital life \cite{mortier2016personal} and acts as a personal locally-stored data repository to empower users to manage their personal data.

Privacy and crowdsourcing are the governing themes of our study, while the IoT space is a fitting area of application. With the databox architecture and data anonymisation in mind, we propose a text masking method for the analysis of social media submissions. Our method transforms text to vectors and/or images, challenging the perception of participating workers. 

\section{Proposed methodology}

\begin{figure}[ht]
\centering
\includegraphics[width=.9\linewidth]{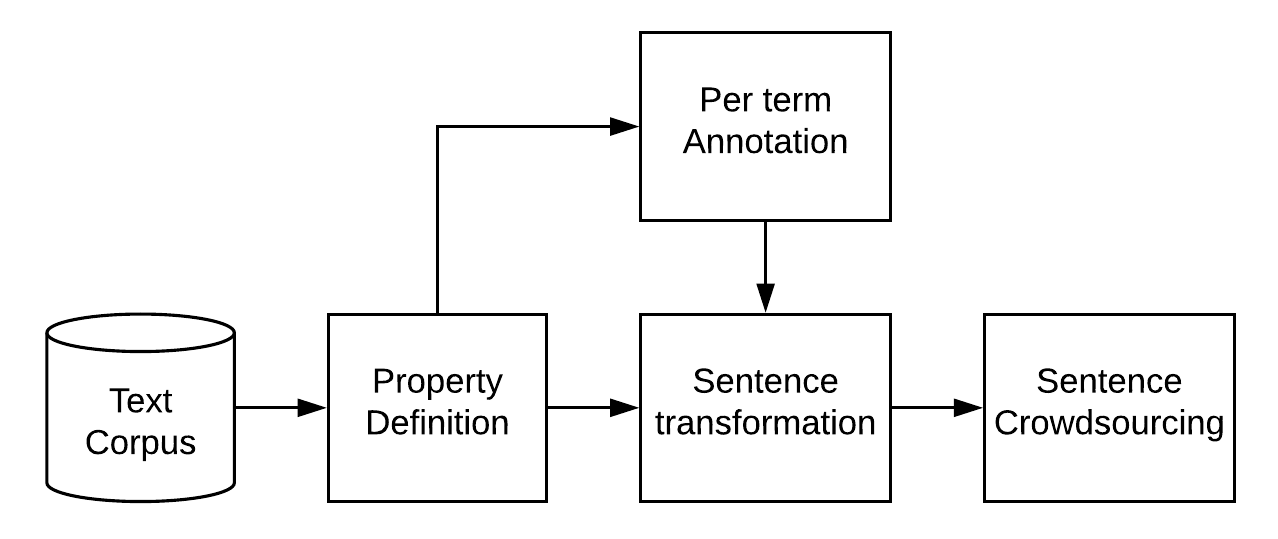}
\caption{The process of text transformation}
\label{process}
\end{figure}

Our proposed process of text transformation, Figure \ref{process}, can be summarised as follows. Given a preprocessed cleaned text corpus, we define the text properties of interest. We then use a per term annotation of the desired properties to transform each sentence in the corpus via crowdsourcing. Even if the per-term annotation labels do not exist, they can easily be crowdsourced as single terms, with no privacy concerns.

\subsection{Lexicon}

The sentiment of text can be defined through human labelling. In OSNs, the text collection could be comprised of submissions made from users. The original submission is shown to annotators, who classify the submission according to its sentiment(s). Although public posts in OSNs are considered as public domain, OSN users do not provide an explicit consent for the use of their data in a labelling task, while annotators can easily trace the original author via simple search engine queries.

We propose a privacy preserving transformation, where words are replaced by their properties. In sentiment analysis applications, the property of interest is emotion. In our study, each word is represented by the emotion it conveys. The Pure Emotion Lexicon (PEL) \cite{haralabopoulos2018multivalued,haralabopoulos2017crowdsourcing} contains a beyond polarity emotion vector, instead of a single emotion \cite{deng2015mpqa,strapparava2004wordnet}. The emotional vectors are normalized emotion classification results for each term and correspond to the eight basic emotions \cite{plutchik1980general}.

For instance, the word "normal" received the following annotations in PEL:

\begin{center}
\small
\begin{tabular}{c c c c}
anticipation&0 & sadness&0\\
joy&3 & disgust&0\\
trust&4 & anger&0\\
fear&0 &surprise&0\\
\end{tabular}
\end{center}

\noindent Its emotional vector is:\\
\centerline{[ 0, 3, 4, 0, 0, 0, 0, 0 ]}
\\
\noindent and its normalized vector is:\\
\centerline{[ 0, 0.75, 1, 0, 0, 0, 0, 0 ]}
\\

\begin{figure}[ht]
\centering
\includegraphics[width=.9\linewidth]{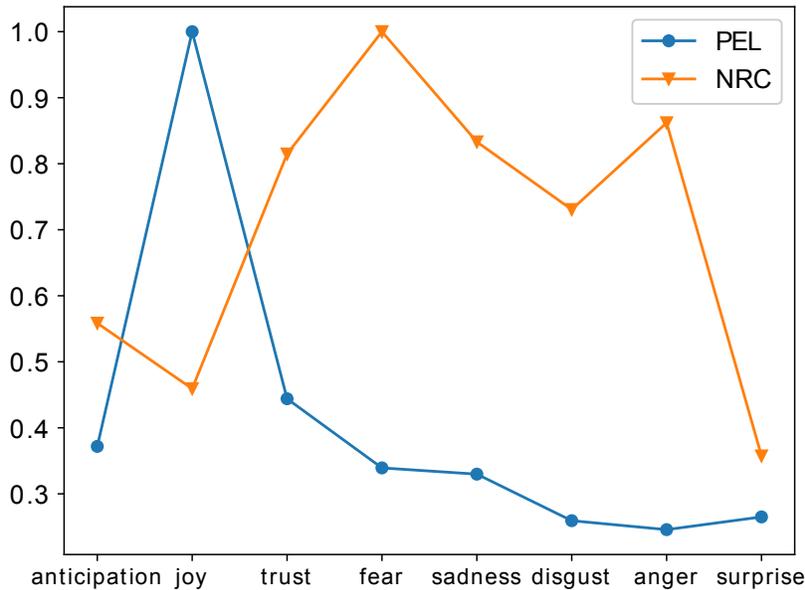}
\caption{Distribution of emotions in lexicons}
\label{fig:lexdist}
\end{figure}

The NRC lexicon \cite{Mohammad13,mohammad2010emotions} is also converted to the same normalized vector format of the eight basic emotions. In total, PEL\footnote{\url{https://github.com/GiannisHaralabopoulos/Lexicon}} contains 9736 stems from 17739 terms, while NRC\footnote{\url{http://www.saifmohammad.com/WebPages/NRC-Emotion-Lexicon.htm}} included 3860 stems based on 4463 terms. The emotional distribution within each lexicon can be seen in Figure \ref{fig:lexdist}. PEL is dominated by joy annotations, while NRC has a high number of fear annotations.

\subsection{Privacy}

The mathematics formulation would be as following. Let $d$ be a text collection with $n$ number of words $w$. 

\begin{equation}
d=[w_1,w_2,....,w_n]
\end{equation}

A word $w_i$ is a vector of 8 elements, representing the properties of each word, $e\,\epsilon\,[0,1]$:

\begin{equation}
w_i=[e_1,e_2,....,e_8]
\label{eq:word}
\end{equation}

Assuming element $e$ can only have two decimals (i.e. $e$ can have one out of $101$ possible values), the number of possible vector permutations for a word $w$ is $101^8$.

A document $d$ with $n$ number of words has:

\begin{equation}
(101^8)^n
\end{equation}

possible permutations. The number of possible three words sentences is more than $10^{48}$. In comparison, a 256-bit encryption method has roughly $10^{77}$ different keys.

\begin{figure}[ht]
\includegraphics[width=\linewidth]{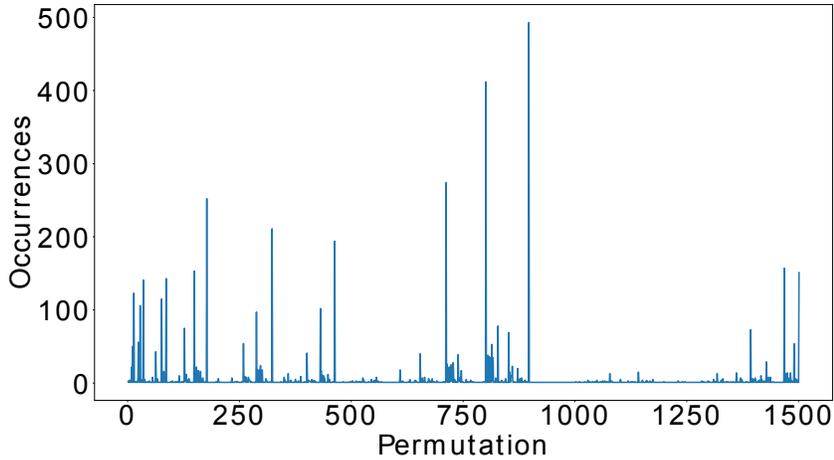}
\caption{Distribution of permutations in lexicon}
\label{perm}
\end{figure}

Currently, there are 1502 different emotion vector permutations in the PEL lexicon, distributed as shown in Fig \ref{perm}. 

Given that the emotional vectors are unknown, the permutations for a document $d$ with $n$ number of words from PEL lexicon is:

\begin{equation}
(1502)^n
\end{equation}

The number of possible three word sentences is almost $10^{10}$. 


As the number of possible word permutations increases, the identification of the post (and/or the user that submitted it) becomes more and more complex. So that with only three words in a sentence, the number of possible three words sentences is enough to guarantee a high level of privacy. This is without taking into account the image transformation variability, e.g. colour hues, which will be presented in the following subsection.


\begin{center}
\centering
\small
\begin{tabular}{cccccccc}
ant & joy &tru &fear &sad &dis &ang &sur\\
0&0&0&0&0&0&0&0\\
0&0&0&0&0&0&0.25&0\\
0&0&0&0.33&0.33&0.33&0&0\\ 
0&0.15&0&0.85&0&0&0&0\\ 
\end{tabular}
\captionof{table}{List of Vectors Transformation} \label{tab:lovt}
\end{center}

\subsection{Transformation}

The challenge is to create a representation of text, based on word-property association (emotions in our case) vectors, that will retain labelling performance and annotation diversity. To that end we propose two document transformations that preserve privacy. The proposed text transformations are: List of Vectors (LoV) and Image Vectors (IV).

\begin{figure}[ht]
\centering
\includegraphics[width=.6\linewidth]{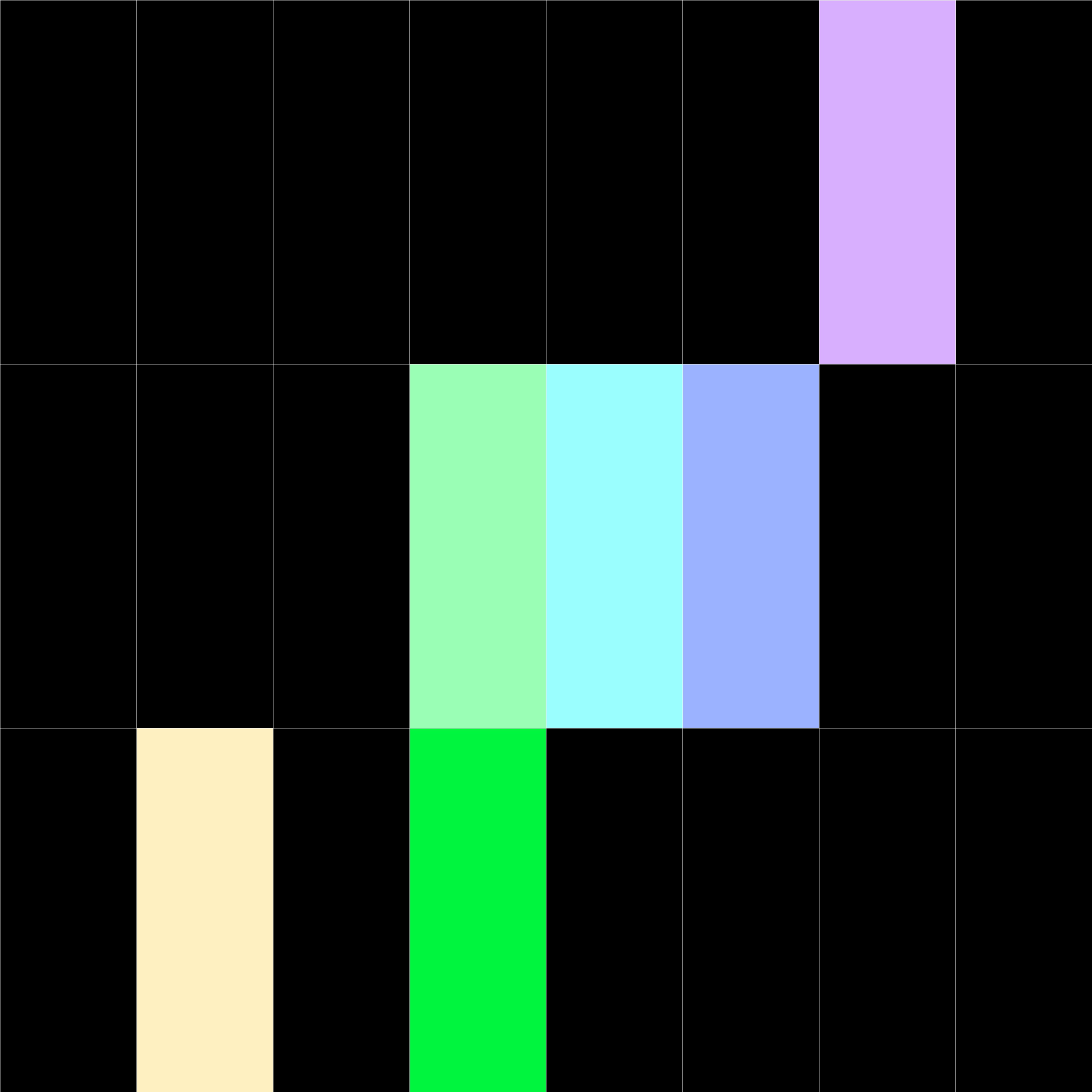}
\caption{Image Vector Transformation}
\label{fig:ivt}
\end{figure}

\begin{figure}[ht]
\centering
\includegraphics[width=.6\linewidth]{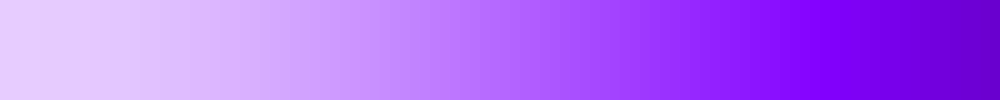}
\caption{Anger hue range based on vector value}
\label{fig:anger}
\end{figure}

LoV and IoV transformations rely on the vector representation of each word, Equation \ref{eq:word}. Let us use the sentence \textit{"They have corruption issues"} to demonstrate the transformation for each method. By using the sentiment vectors we can create an ordered list of representation vectors for LoV, Table \ref{tab:lovt}, where normalized row values correspond to the eight basic emotions (anticipation, joy, trust, fear, sadness, disgust, anger, surprise), as defined by Plutchik in \cite{plutchik1980general}.

Image Vector Transformation (IV) uses the non zero rows of the LoV transformation to create an image representation of the sentence, Figure \ref{fig:ivt}. Words without at least one emotional annotation, (i.e. "the"), are not drawn. Each vector value is transformed into a certain RGB colour with variable hue. The hue is exponentially analogous to the value it represents, the full hue range for anger can be seen in Figure \ref{fig:anger}.

The aforementioned example deals with a part of a sentence. Given a document, the analysed parts/sentences can be aggregated, e.g. based on emotional valence, to provide the overall document sentiment. In our study this aggregation is the averaged sum of normalised emotion values per term in a single sentence and its calculated difference to the mean lexicon emotion. 

Four crowdsourcing tasks are performed, per lexicon. In each task, annotators are presented with only one transformation. E.g. The third task would present LoV tables to annotators, and they had to decide on the dominant emotion based on matrices similar to Table \ref{tab:lovt} with no knowledge of the underlying terms.

\begin{center}
\centering
\small
\begin{tabular}{ c c c }
No Privacy & \textbf{Text}\\
Low Privacy & \textbf{Shuffled} \\
Medium Privacy & \textbf{List of Vectors (LoV)}\\
High Privacy & \textbf{Image Vectors (IV)}\\
\end{tabular}
\captionof{table}{Privacy Levels}\label{tab:priv}
\end{center}

\section{Experiments}

We consider 4 privacy levels in correspondence with the text transformations described above, Table \ref{tab:priv}. Given an online text  submission, the complexity of identifying the user -that submitted the information- increases with each privacy level. We created 4 crowdsourcing tasks per lexicon to analyze the labelling performance of the crowd in diverse privacy settings.

The crowdsourcing tasks were hosted in FigureEight\footnote{\url{https://www.figure-eight.com/}} crowdsourcing platform. The platform was chosen due to a great number of contributor filtering options. We selected contributors with higher than B.Sc. education, are native English speakers and have the highest level of task completion in the platform. We assess the quality of each participant in our tasks with a subjective quality assurance method that inject objective sentences into the subjective corpus \cite{haralabopoulos-etal-2020-objective}. These injected sentences are then used to define a metric of confidence for each participant. The contributions of participants with 90\% or lower confidence were discarded. We also apply a spamming filter at 30\% single annotation percentage on all sentence annotations except LoV transformation for "book" and "osn" sources, where 40\% and 45\% thresholds were applied to retain a sample of statistical significance.

Each sentence in each of the tasks received exactly 10 annotations. Contributors were able to only contribute in one of the tasks, and were excluded from the other three tasks. The use of external crowdsourcing eliminates biases that exist in an internal crowdsourcing task \cite{haralabopoulos2019paid}. We ask contributors a simple question "What is the dominant emotion/colour?". Participants in Text had the full text presented to them, while in all other tasks participants could only view the corresponding transformation and not the initial text. In all tasks except IoV transformation task, the available answers are the eight basic emotions as defined by Plutchik \cite{plutchik1980general}. In the IoV transformation task, the available answers are eight colours, based in the circumplex of emotions \cite{plutchik1980general}.

Each set consists of 100 sentences with terms contained in both PEL and NRC lexicons. The sentences were obtained from three sources, a book\footnote{\url{https://www.gutenberg.org/ebooks/135}} a news site\footnote{\url{https://open-platform.theguardian.com/}}, crawled from Reddit\footnote{\url{https://www.reddit.com/}} and Twitter\footnote{\url{https://twitter.com/}}. These sources will provide diversity in both formality, sentence size and vocabulary \cite{kiritchenko2014sentiment}.

The analysis is performed at a higher level without looking at each separate sentence. As the labels and term annotations are provided via crowdsourcing by anonymous contributors and the task is purely subjective, we only assess contributors based on a confidence level \cite{haralabopoulos-etal-2020-objective}. The results are analysed in the triad of Distribution, Difference and Dominance.

Each of the four privacy levels requires different thought process during labelling. \textit{No} and \textit{Low Privacy} levels provide contributors the text without any emotional information, with \textit{Low Privacy} shuffling the words randomly. \textit{Medium privacy} provides numeric values that correspond to the emotional significance of each term, while \textit{High privacy} level only provides a palette of colours where the emotional significance is represented by hue. 

As mentioned, emotional labelling is mainly subjective. Still, our goal is to assess whether these transformations of text can provide similar results to traditional text labelling. Thus, we will use \textit{No Privacy} labels as baseline for comparison with the other Privacy levels. However, \textit{No Privacy} labels should not be interpreted as the correct labels nor as the gold standard. We will present two sets of results, based on the lexicons used to create the transformations, PEL or NRC.

\begin{figure}
\centering
\includegraphics[width=.9\linewidth]{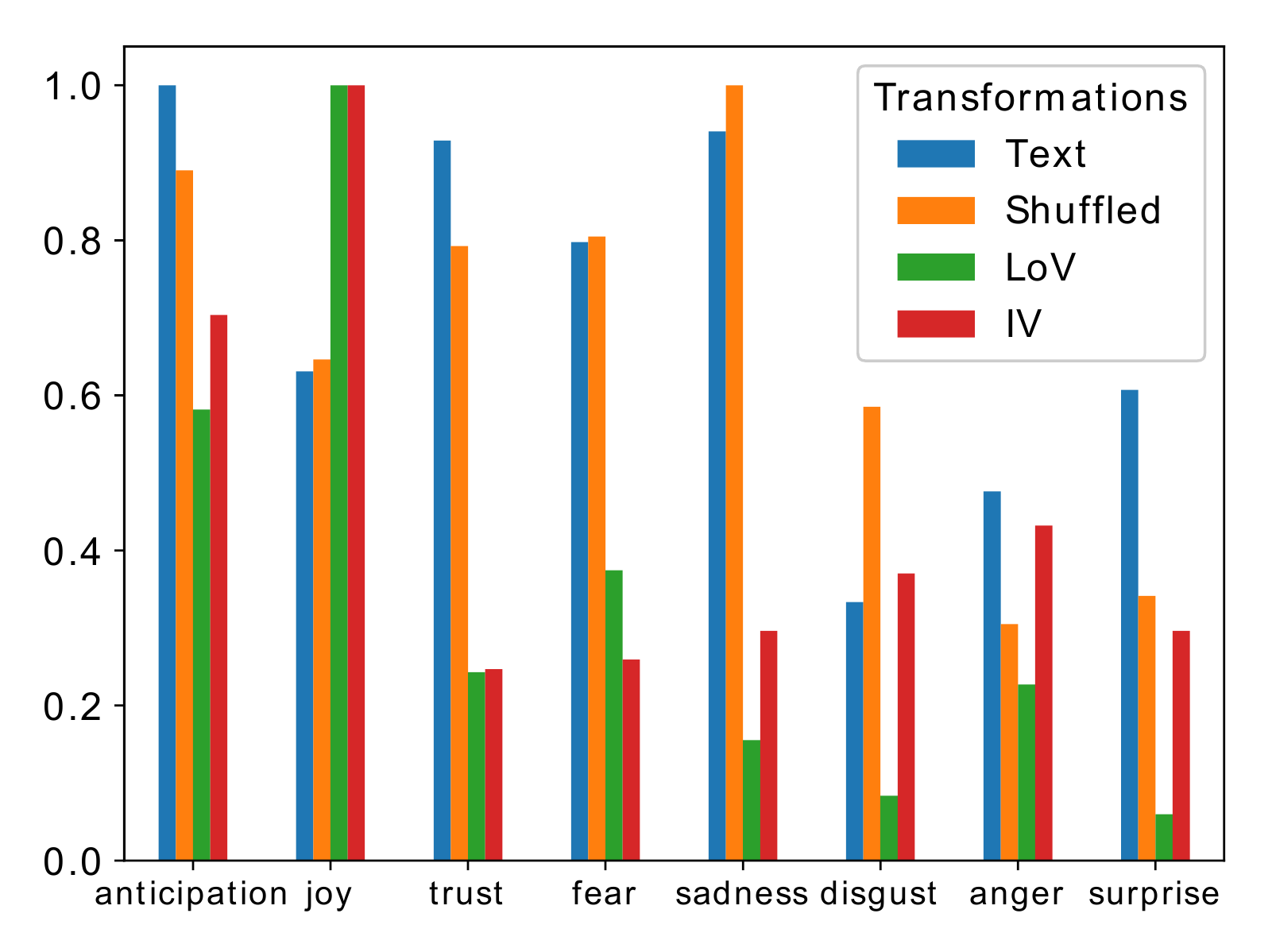}
\caption{Distribution of annotations, \textit{book} source}
\label{fig:sentannbook}
\end{figure}

\subsection{PEL}

\begin{figure}
\centering
\includegraphics[width=.9\linewidth]{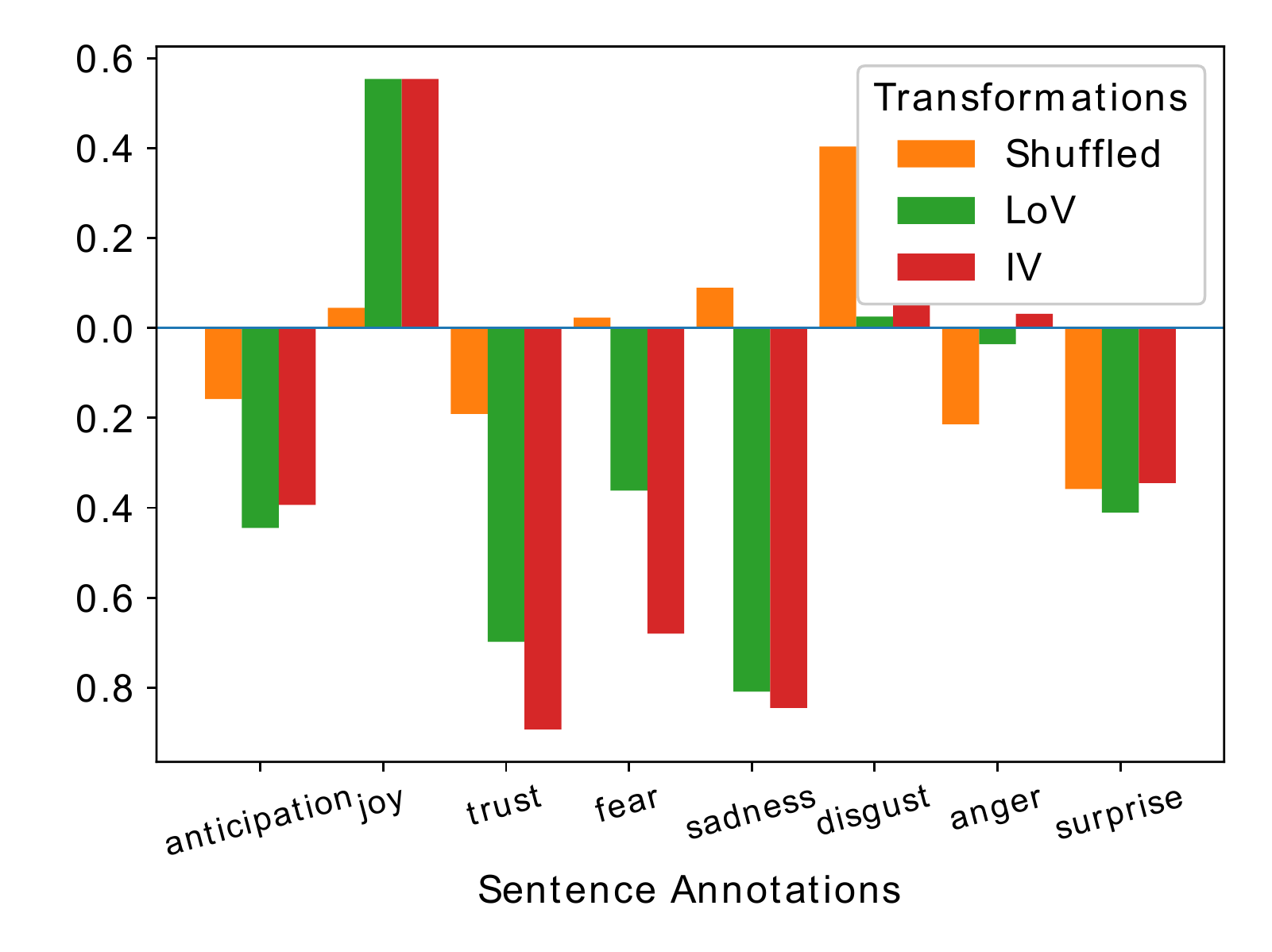}
\caption{Annotational differences to Text, \textit{book} source}
\label{fig:anndifftextbook}
\end{figure}

The distribution of sentence annotations, Figure \ref{fig:sentannbook} is compared in pairs. The first pair includes the methods that do not provide any term specific emotion information, Text and Shuffled, while the second couple consists of the methods that provide a transformed emotional information, LoV and IV. 

Within each pair, the annotations don't vary more than 20\%. LoV and IV are created by using lexicon annotation distributions, therefore PEL acts as the transformation agent of a sentence. Thus annotations are bound to PEL distribution, Figure \ref{fig:lexdist}. Regardless of the source of sentences, joy sentence annotations are prominent in LoV and IV transformations. Labels obtained through IV provided an annotation distribution closer to Text than LoV in 19 out of 24 occasions (3 sources with 8 emotions each), and in fewer cases outperformed Shuffled.

\begin{figure}
\centering
\includegraphics[width=.9\linewidth]{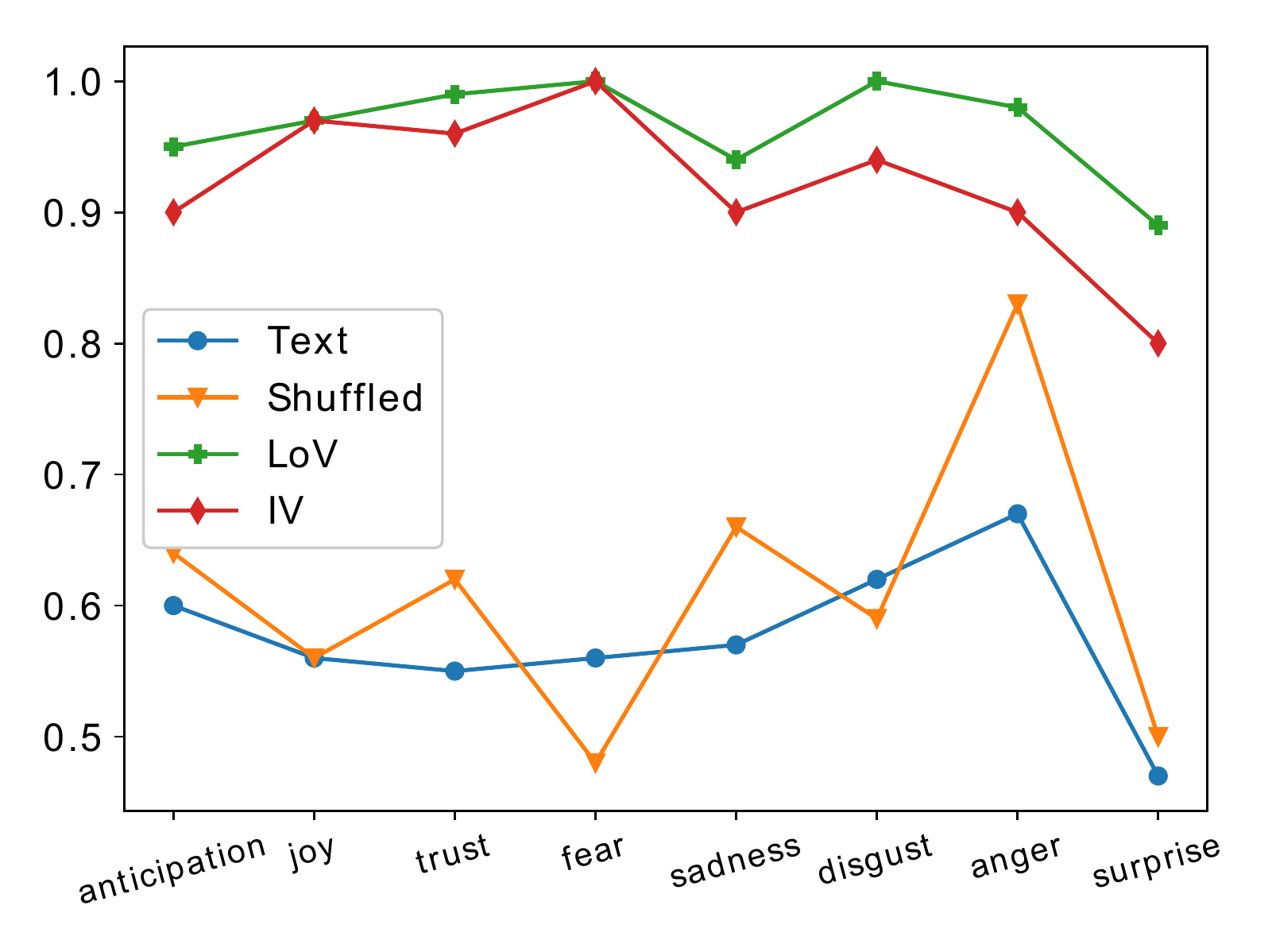}
\caption{Dominant emotion agreement for sentences, \textit{book} source}
\label{fig:domemotionbook}
\end{figure}

The annotational difference chart, shown in Figure \ref{fig:anndifftextbook}, further highlights the high number of joy annotations in the PEL lexicon, reflected in both LoV and IV transformations. Sentence annotations via Shuffled method present a slight variation in four out of eight emotions. A positive variation of 'joy' emotion annotations exists in sentences from all sources.

Sentences from \textit{book} source have 'fear', 'trust' and 'sadness' present the highest mean negative annotation difference, compared to Text sentence annotations. \textit{News} source sentences have the highest positive difference and the lowest mean negative difference. The emotions: 'fear', 'sadness', 'anger', 'surprise' are negatively affected by LoV and IV transformations. While, sentences from \textit{osn} source have significant positive 'joy' and negative 'trust' differences due to LoV and IV transformations.

\begin{table}
\small
\centering
\begin{tabular}{c | c | c | c }
Ant. & Joy: 0.94 & \textbf{Ant: 0.62} & Tru: 0.51 \\
\hline
Joy&\textbf{Joy: 1.89} & Ant: 0.58 & Tru: 0.5\\
\hline
Trust & Joy: 0.37 & \textbf{Tru: 0.36} & Dis: 0.12\\
\hline
Fear & Joy: 0.18 & Ant: 0.13 & \textbf{Fear: 0.13} \\
\hline
Sadness& Ant: 0.17 & Joy: 0.14 & Tru: 0.11 \\
\hline
Disgust & Joy: 0.29 & \textbf{Dis: 0.12} & Tru: 0.12 \\
\hline
Anger & Tru: 0.05 & Joy: 0.04 & Ant: 0.03\\
\hline
Surprise & \textbf{Sur: 0.32} & Joy: 0.22 & Tru: 0.19\\
\end{tabular} 
\caption{Sentence annotation \textit{(column 1)} and top three IV term aggregation scores \textit{(columns 2-4)} for PEL lexicon, \textit{osn} source}
\label{tab:PELagg}
\end{table}

 The strength of the dominant emotion for each sentence label, as defined by the majority strength of the annotations received, is portrayed in Figures \ref{fig:domemotionbook}. The Text and Shuffled transformation have a low dominant emotion agreement, which reflects the diversity of opinions, the subjective nature of the annotation task. Transformations of LoV and IV have high dominant emotional agreement, most probably due to differences in the presentation of the task. Both methods "guide" contributor with objective visual elements, number and colours, while the subjectivity of the task is transfered to the comparison of each -objective- element.

 \begin{figure}
\centering
\includegraphics[width=.9\linewidth]{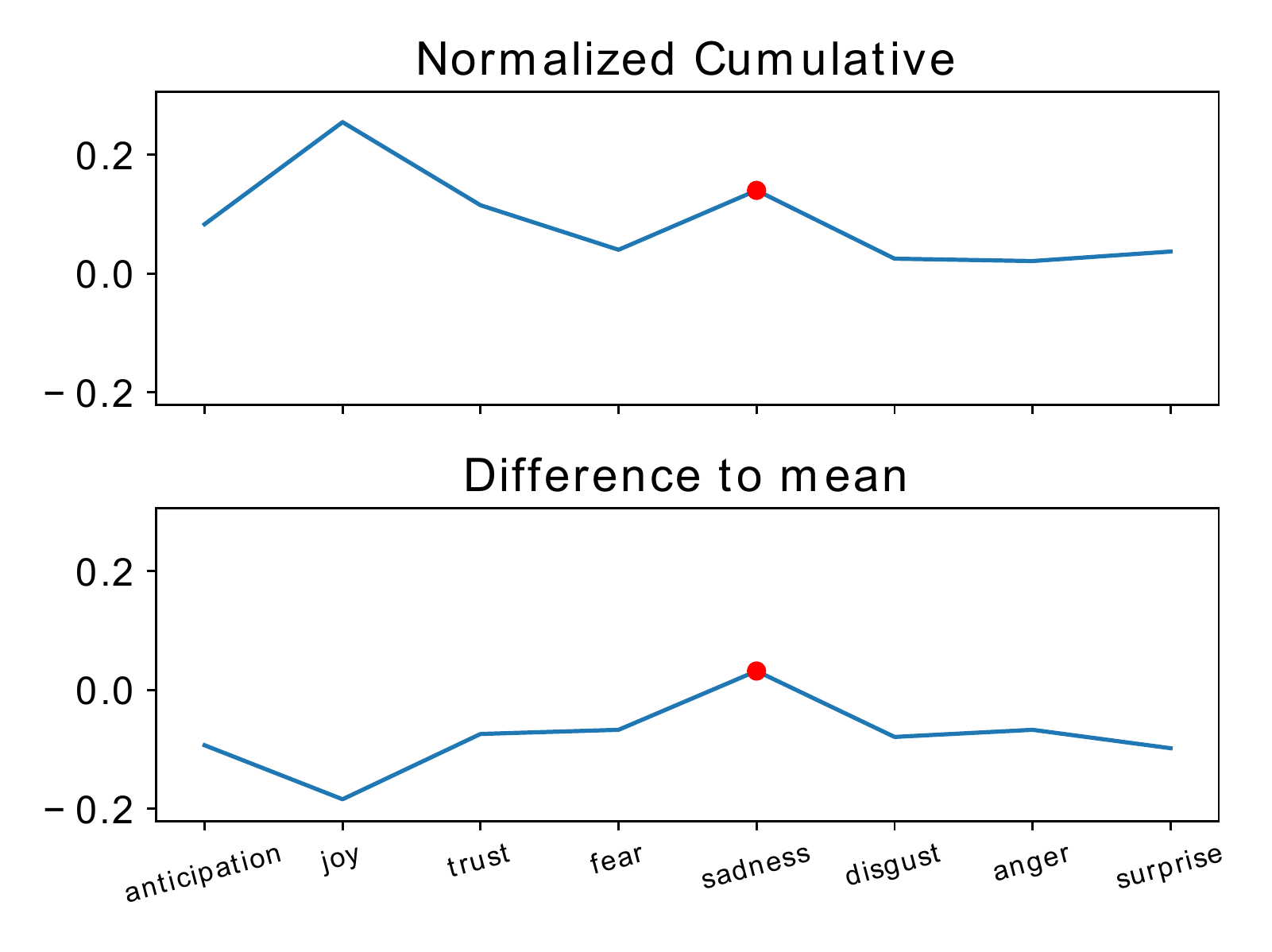}
\caption{Per term aggregation based on PEL and Difference to mean PEL aggregation for \textbf{sadness} sentences via \textit{IV}}
\label{fig:BookIVsadnessPEL}
\end{figure}

 Dominant emotion is characterised, on all sources, by low Text and high LoV and IV agreement, with Shuffled sentence agreement varying across sources. Sentences from \textit{book} are a good example of that, Fig. \ref{fig:domemotionbook}. Sentences from \textit{news} source have similarly low Text and high LoV agreement, but Shuffled and IV transformations fall within the area of 65\% to 90\%. The absence of 'surprise' dominant emotion agreement is due to the absence of sentences majorly annotated with 'surprise' emotion. Regardless of the source, LoV presents the highest level of agreement followed by IV.

When we aggregate the per term emotion vectors of a sentence, joy is more prominent. However, when we calculate the difference of the normalized cumulative sentence emotion to the normalised mean lexicon emotion, we can uncover the prominent sentence emotion with high enough certainty, Figure \ref{fig:BookIVsadnessPEL}. A correlation of our difference aggregation and the actual sentence annotation is evident in most transformations and sources. The cumulative emotion of the terms in each sentence is not majorly related to its annotation as a sentence, due to the high number of 'joy' annotations in the PEL lexicon. However, even with the single emotion skewed PEL lexicon, the sentence label can be approximated by calculating the mean absolute difference of the exact emotion annotations to the average source annotations.

\begin{figure}[ht]
\centering
\includegraphics[width=.9\linewidth]{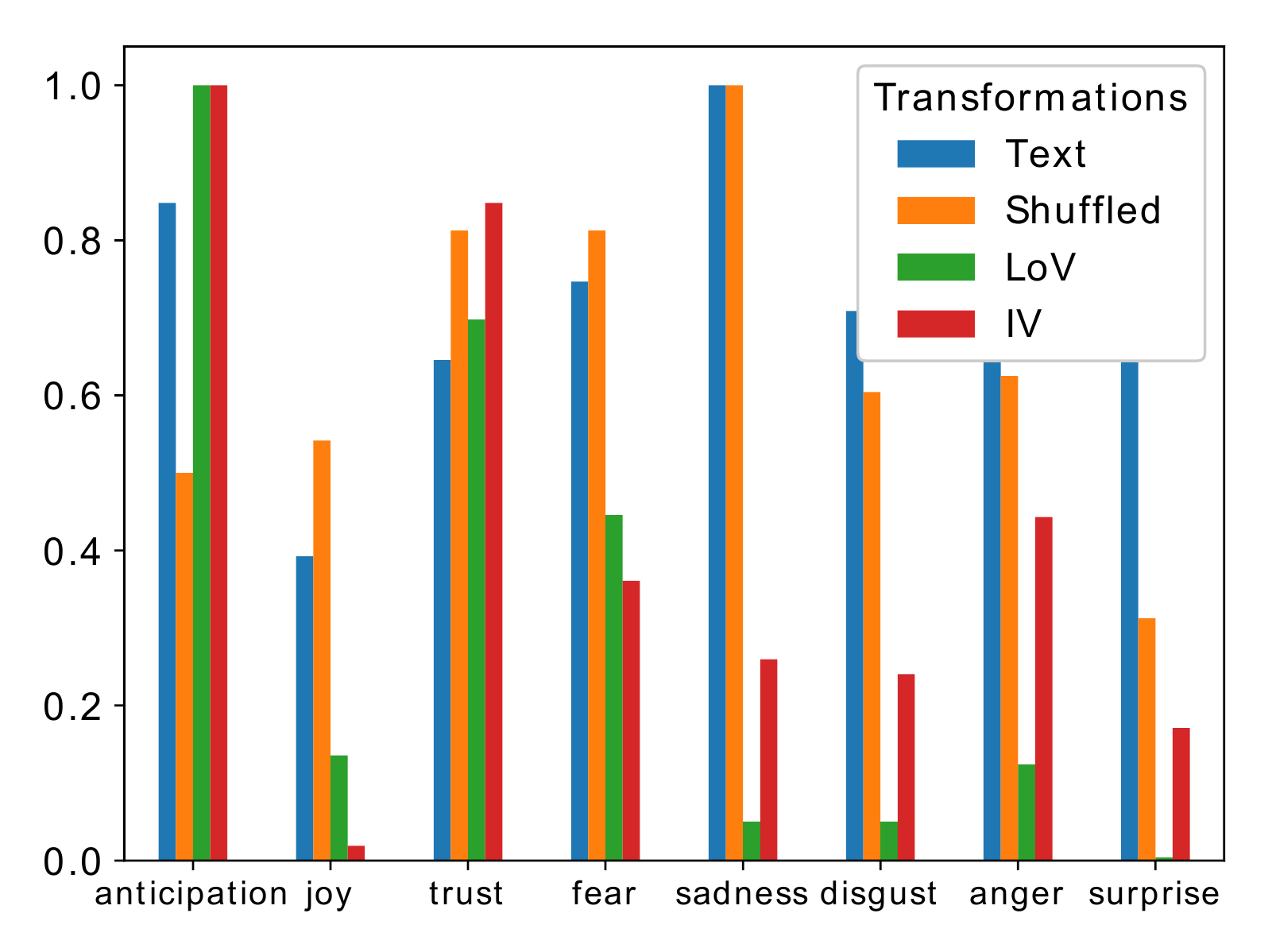}
\caption{Distribution of annotations, \textit{news} source}
\label{fig:sentannnewsNRC}
\end{figure}

Table \ref{tab:PELagg} presents the simple aggregation of terms in relation to the emotion annotation of the sentence for \textit{osn}. On the IV transformation and based on PEL term emotion vectors aggregation, six out of eight emotions appear in the top three aggregation scores. In total, a degree of correlation is present in more than 70\% of the 72 cases, 8 emotions with 4 transformations and 3 sources. The simple aggregation of terms presents high joy annotations, especially in LoV and IV transformations, Table \ref{tab:PELagg}. Per term simple aggregation of emotions 'sadness' and 'anticipation' is in accordance with sentence annotations in almost half of the cases. For sentences annotated without any transformation, text and shuffled, the aggregation of per term emotions corresponds to the sentence annotation in four out of eight emotions, in two cases it is the second highest, and in the rest two the third highest emotion. 

Regardless of the transformation and the source of sentences, the distribution difference corresponds to the base emotional annotation in $96\%$ of IV transformations. A strong indication of term to sentence relationship, as mentioned in \cite{socher2013recursive,giatsoglou2017sentiment}. 

\subsection{NRC}

\begin{figure}
\centering
\includegraphics[width=.9\linewidth]{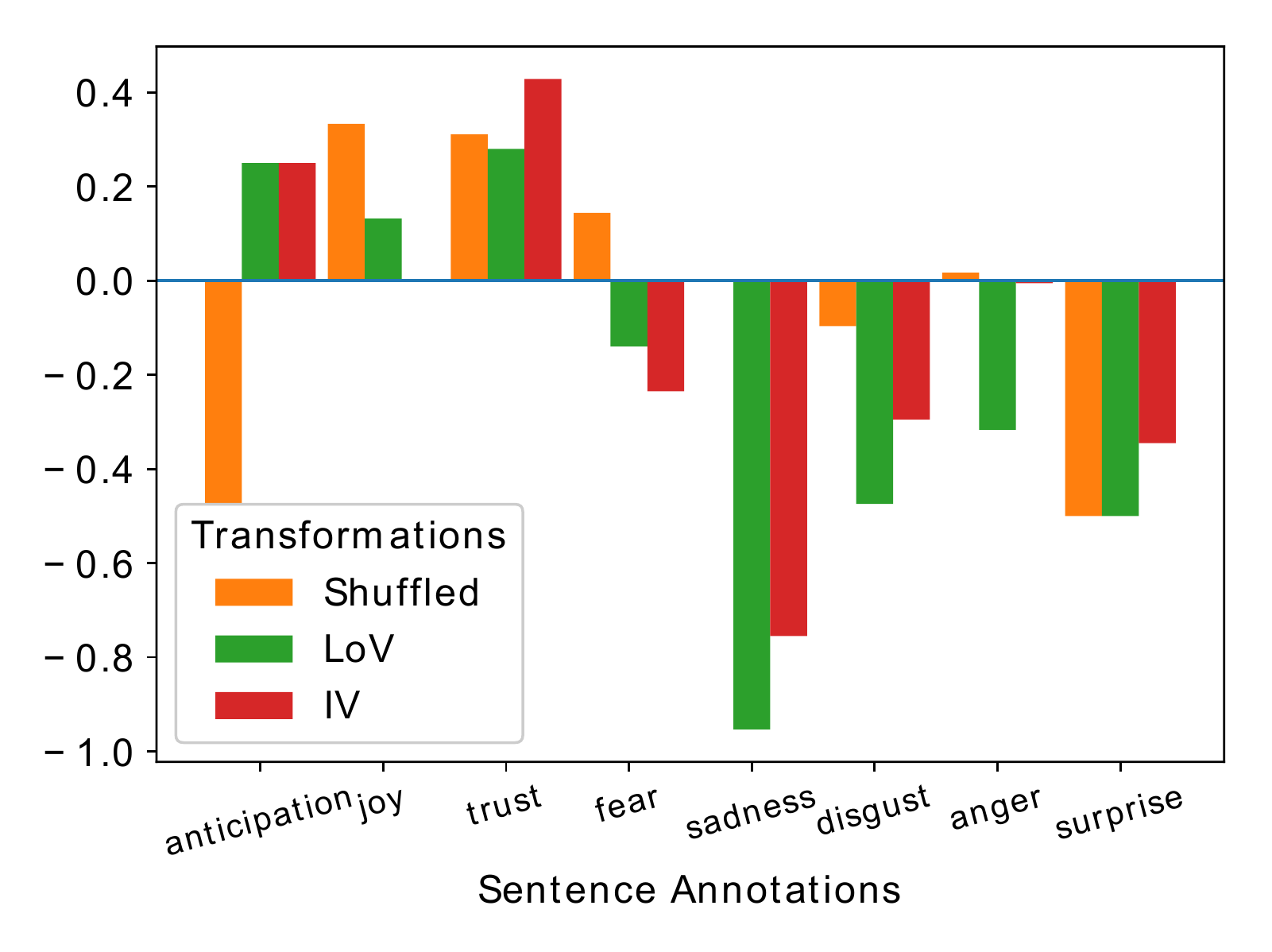}
\caption{Annotational differences to Text, \textit{news} source}
\label{fig:anndifftextnewsNRC}
\label{figset:diffNRC}
\end{figure}

The distribution of sentence annotations for \textit{news} source with NRC transformation is portrayed in Figure \ref{fig:sentannnewsNRC}. Annotations for LoV and IV present a significant difference compared to PEL. IV annotations difference to Text is lower than LoV difference in $58.3\%$ of the cases. Regardless of the source, 'anticipation' is the most frequent sentence label, followed by 'trust'. For sentences from \textit{book} source, four out of the eight emotions received low number of annotations on IV, but sadness received significantly more on LoV. Sentences from \textit{osn} had the most balanced annotations and the highest correlation of IV to Text annotations. Joy was the least annotated emotion in \textit{news} for IV transformation, which is in line with the low number of joy annotations in text transformation.

\begin{figure}
\centering
\includegraphics[width=.9\linewidth]{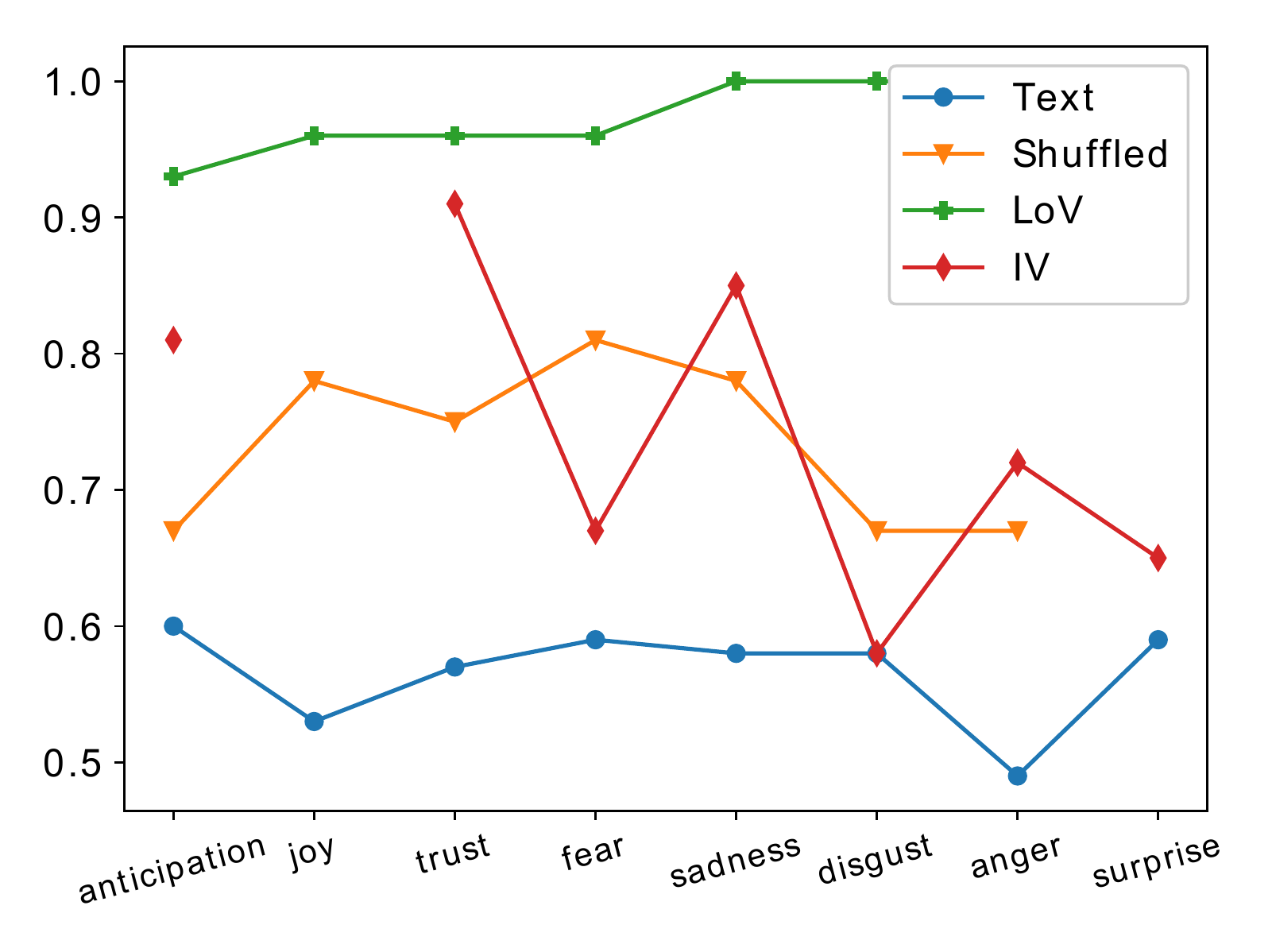}
\caption{Dominant emotion agreement for sentences, \textit{news} source}
\label{fig:domemotionnewsNRC}
\end{figure}

The mean annotational difference of LoV and IV to Text is lower than the one observed in PEL. Sadness in \textit{news} source has the highest difference to the annotations of Text, Figure \ref{fig:anndifftextnewsNRC}. In some cases shuffled annotations have higher difference than LoV and/or IV annotations, but IV annotations have a greater negative difference in \textit{news} sources.

Regardless of the source, there is a clear reduction in emotional diversity, with one or two emotions receiving higher number of annotations, while the rest of the emotions are negatively affected. A similar performance was observed in PEL annotational differences. In PEL 'joy' was a key factor in transformation, whereas NRC emotions that positively influence annotations are 'anticipation', 'trust' and 'joy'.

\begin{figure}
\centering
\includegraphics[width=.9\linewidth]{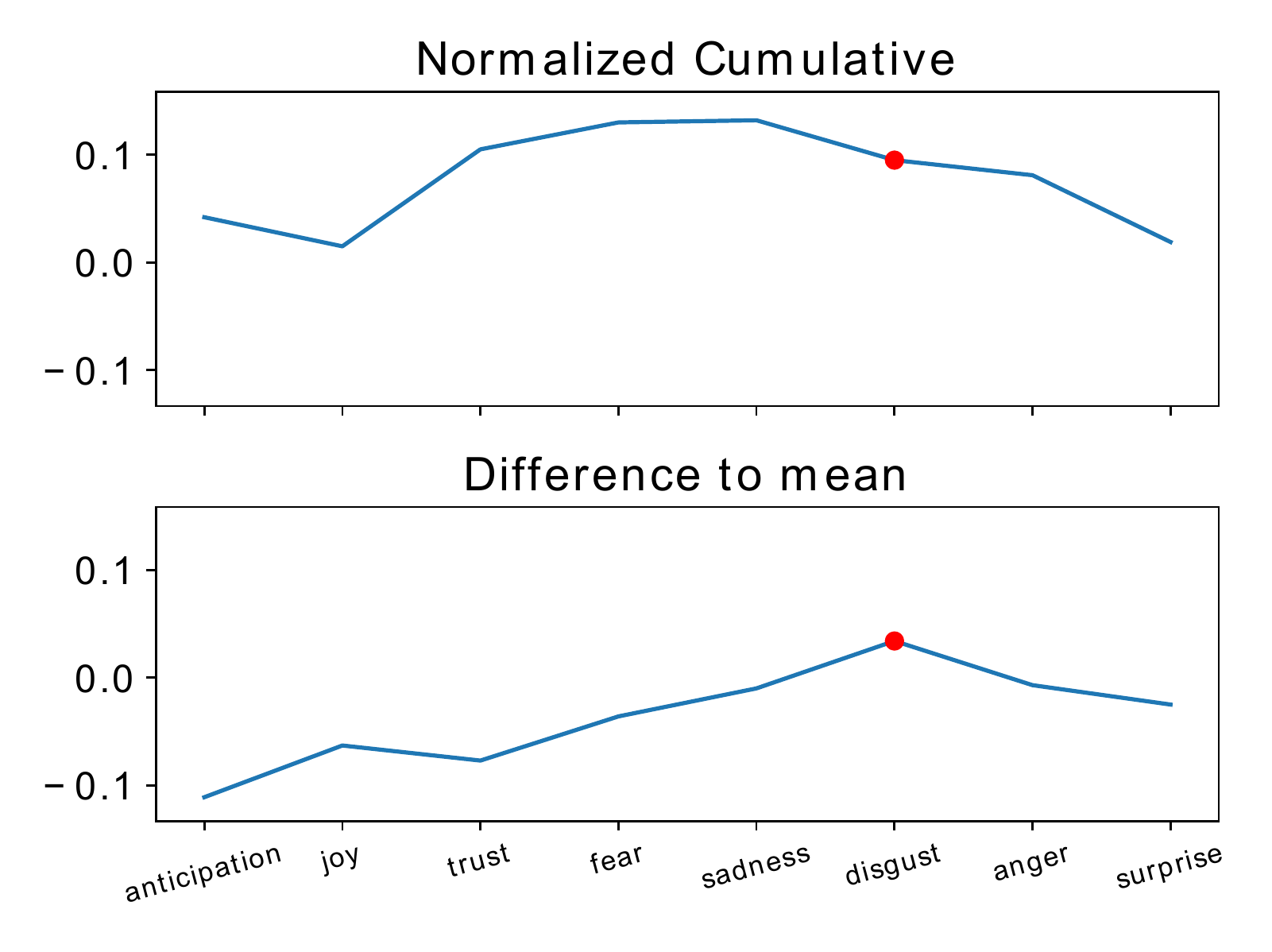}
\caption{}
\label{fig:NewsIVdisgustNRC}
\caption{Per term aggregation based on NRC and Difference to mean NRC aggregation for sentences annotated with \textbf{disgust} via \textit{IV}}
\end{figure}

The annotational agreement for sentences from \textit{book} sources is similar to the one observed in PEL transformed sentences. Dominant emotion agreement is low on Text, variable on Shuffle, and high for LoV and IV. The rest of the emotions present the same agreement as in \textit{book}. On the contrary, the emotion agreement in \textit{news} suggests a greater diversity of opinions, Figure \ref{fig:domemotionnewsNRC}. Agreement is low on Text and high on LoV, but Shuffled and IV intertwine in the space between the two. As text labelling is an subjective task, we seek as much diversity as possible.

\begin{table}

\centering
\small
\begin{tabular}{c | c | c | c }
Ant. & \textbf{Ant: 0.52} & Joy: 0.3 & Tru: 0.27\\
\hline
Joy & \textbf{Joy: 0.15} & Tru: 0.11 & Ant: 0.06\\
\hline
Trust & \textbf{Tru: 0.6} & Ang: 0.25 & Ant: 0.23 \\
\hline
Fear & Tru: 0.22 & \textbf{Fear: 0.18} & Ang: 0.12 \\
\hline
Sadness & \textbf{ Sad: 0.23} & Ant: 0.15 & Fear: 0.12 \\
\hline
Disgust & Tru: 0.11 & Joy: 0.05 & Ant: 0.04 \\
\hline
Anger & Fear: 0.14 & Tru: 0.11 & \textbf{Ang: 0.11} \\
\hline
Surprise & \textbf{Sur: 0.08} & Joy: 0.07 & Tru: 0.07\\
\end{tabular} 
\caption{Sentence annotation \textit{(column 1)} and top three IV term aggregation scores \textit{(columns 2-4)} for NRC lexicon, \textit{osn} source}
\label{tab:NRCagg}
\end{table}

Throughout the transformations, the high number of 'trust' annotations in NRC, results in high aggregations for the majority of sentence labels, with a comparatively low aggregation score for 'disgust', 'anger' and 'surprise' Table \ref{tab:NRCagg}. Five out of eight emotions for \textit{osn} source with IV transformation and based on NRC emotion aggregation are highly correlated with the sentence annotation, Table \ref{tab:NRCagg}, and the difference between the highest emotion aggregations are significantly smaller than in PEL. More than $50\%$ of the sentence emotions, for all 72 cases of IV transformation, can be accurately predicted based on a per term aggregation. Similar to PEL, per-emotion annotation difference to the mean lexicon emotional distribution can also be used to determine the most appropriate emotion in more than $93\%$ of the cases, Figures \ref{fig:NewsIVdisgustNRC}.

\begin{table}[ht]
{
\centering
\small
\begin{tabular}{c | c | c | c }
\textbf{Source} & \textbf{Method} & \textbf{PEL} & \textbf{NRC}\\\hline
\multirow{ 3}{*}{Book} & Shuffled & \multicolumn{2}{c}{ 0.2645}\\
 & LoV  & 0.0678 & 0.2065\\
 & IV  & 0.0615 & 0.1522\\\hline
\multirow{ 3}{*}{News} & Shuffled & \multicolumn{2}{c}{ 0.1788}\\
 & LoV  & -0.0001 & 0.0921\\
 & IV  & 0.0069 & 0.0988\\\hline
\multirow{ 3}{*}{Osn} & Shuffled & \multicolumn{2}{c}{ 0.1677}\\
 & LoV  & 0.0611 & 0.1365\\
 & IV  & 0.0567 & 0.0905\\
\end{tabular} 
\caption{Mean Spearman's Rho per source and method, compared to Text annotations (Shuffled is independent of lexicon)}
\label{tab:spearman}
}
\end{table}

\subsection{Correlation to traditional labelling}

Spearman's Rho correlation of annotations from Text to all transformations with PEL is low, but greatly improves when using the NRC lexicon, Table \ref{tab:spearman}. Shuffled was included as a simple privacy measure, but it presents the highest correlation to Text annotations, although it often exhibits sentiment loss due to rearrangement of terms. The use of a higher quality lexicon improves the correlation. This translates to: improving the transformation agent improves the quality of the labelling process.

\section{TF-IDF weighing}

\begin{center}
\centering
\small
\begin{tabular}{cccccccc}
ant & joy &tru &fear &sad &dis &ang &sur\\
0&0&0&0&0&1 (.48) &0&0\\
0&0&0&0&.06 (.32) &0&0&0\\
0&.33 (.21)&0&0&0&0&.33 (.21)&0\\
\end{tabular}
\captionof{table}{LoV transformation before TF-IDF (after TF-IDF)} \label{tab:pretfidf}
\end{center}

To demonstrate the diversity of our approach, we combine emotion embeddings and a more traditional Term Frequency - Inverse Document Frequency (TF-IDF) calculation \cite{ramos2003using}. For each term in the corpus we calculate its maximum TF-IDF value, ranging from 0 to 1. This values is then multiplied to the whole emotional vector.

\begin{figure}[t]
  \centering
  \begin{subfigure}{.3\columnwidth}
    \centering
    \includegraphics[width=\linewidth]{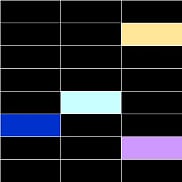}
    \caption{}
  \end{subfigure}%
  \hspace{1em}%
  \begin{subfigure}{.3\columnwidth}
    \centering
    \includegraphics[width=\linewidth]{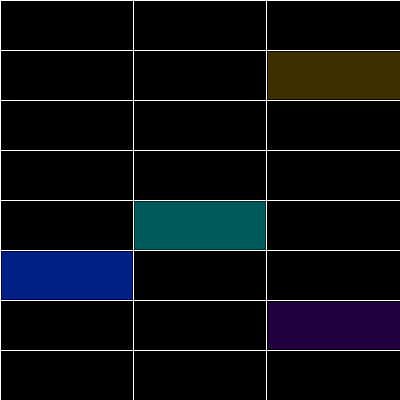}
    \caption{}
  \end{subfigure}%
  \caption{IV transformation before(a) and after(b) TF-IDF}
  \label{fig:tfidfiv}
\end{figure}

For example: the sentence "the insult is not to him but to the law" has an initial LoV transformation as shown in Table \ref{tab:pretfidf} and after TF-IDF is applied, it has an LoV transformation as shown in parenthesis of Table \ref{tab:pretfidf}. Similarly the IV transformation is affected by the TF-IDF weighting. Since each colour is proportionally vibrant to the emotion score, a high TF-IDF enriches colours while a low TF-IDF makes them more subtle. For example "the insult is not to him but to the law" would be transformed as seen in Figure \ref{fig:tfidfiv}. Since none of the terms has a emotional score close to 1, no cell is particularly vibrant in Figure \ref{fig:tfidfiv}b.

We apply the TF-IDF term weighting in the same set of sentence as before. The exact same tasks are performed in the Amazon Mechanical Turk\footnote{\url{https://www.mturk.com/}} crowdsourcing platform. Unfortunately, figure eight platform is no longer accessible to researchers. We therefore decided to perform two additional tasks per source, text and shuffled. Hence the difference between the shuffled values of Table \ref{tab:spearman} and Table \ref{tab:spearmantf} is due to the new tasks performed in a different platform. Nonetheless, this further demonstrates the capabilities of privacy aware crowdsourcing, since the mean shuffled correlation in this experiment is improved by at least 82\% (\textit{book} source) and up to 152\% (\textit{news} source).

 When the PEL lexicon is used as an emotion transformation agent, and in \textit{book} and \textit{news} sources, the IV transformation improves the correlation with the actual sentence, when compared to LoV. This indicates that the visual representation could better convey the emotional information when compared to a list of numerical values. However, similarly to the previous experiments, the NRC functions as a better transformation agent and all the correlations are better than those of PEL. The only occasion where the LoV or IV transformations with TF-IDF are not improving the correlation, when compared to those without TF-IDF, is in \textit{osn}. The most probable explanation for the low correlation values in \textit{osn} is the frequent absence of context from online submissions, which is in turn transferred to the transformations. 

\begin{table}[ht]
{
\centering
\small
\begin{tabular}{c | c | c | c }
\textbf{Source} & \textbf{Method} & \textbf{PEL} & \textbf{NRC}\\\hline
\multirow{ 3}{*}{Book} & Shuffled & \multicolumn{2}{c}{ 0.4815}\\
 & LoV  & 0.141 & 0.2407\\
 & IV  & 0.1535 & 0.1548\\\hline
\multirow{ 3}{*}{News} & Shuffled & \multicolumn{2}{c}{ 0.3848}\\
 & LoV  & 0.0304 & 0.1486\\
 & IV  & 0.1005 & 0.1311\\\hline
\multirow{ 3}{*}{Osn} & Shuffled & \multicolumn{2}{c}{ 0.3609}\\
 & LoV  & 0.0568 & 0.1618\\
 & IV  & 0.0147 & 0.0875\\
\end{tabular} 
\caption{Mean Spearman's Rho per source and method, compared to Text annotations (Shuffled is independent of lexicon) with TF-IDF}
\label{tab:spearmantf}
}
\end{table}

\section{Conclusions}

We presented a novel approach to privacy-aware labelling that retains subjectivity and is performed through crowdsourcing. The key outcome of our study is: the trade-off between privacy and an as-is presentation is interconnected to the trade-off of agreement and diversity. Text transformations that ensures privacy acts as a curb to contribution diversity, a much needed quality in subjective crowdsourcing tasks\cite{haralabopoulos-etal-2020-objective}. We also demonstrated how simple NLP methods, such as TF-IDF weighing, can be used to further improve the correlation results of the LoV and IV transformations. 

We also presented a naive per term emotion aggregation, capable of acting as a method for label validation. The evaluation of the results is performed via a direct comparison of the privacy aware annotations to non private annotations. We refrain from evaluating the labels in a downstream task, as such an evaluation would add a range of new variables. Although manual labelling is not state of the art for NLP machine learning tasks, labelling of sentences and text is widely used in computer applications.

The lexicons used contain a low number of emotional permutations and high level of certain emotion annotations, 'joy' for PEL and 'trust' for NRC. Sentences were split based on punctuation, but different splitting methods (e.g. syntactic) should be studied in order to determine the most appropriate. While the transformation of negation, in a similar text to image scenario, has to be considered. Finally, scaling factor for hues is exponential in our experiment, but different scaling functions can be used to better convey the transformed emotion.

Our proposed text transformation can be combined with data augmentation methods \cite{haralabopoulos2021text} and its application is not limited to sentiment analysis tasks. Psychology researchers are often put up against disturbing reports that affect their well-being \cite{bernal2010itinerant}. A text transformation method can be applied to perform tasks that do not require meticulous study, e.g. a classification of texts based on abuse type. The transition from the traditional text annotation to a more objective visual representation poses challenges to annotators and requesters. Annotators have to adjust their skills to a visual representations, while requesters need to carefully design the transformations in order to preserve the subjectivity of annotations. Emotion transformation is just one of the possible text to property associations that can be used to analyse text. LoV and most importantly IV mask text in a way that provides privacy to the creator and usability to researchers.

\bibliographystyle{plain}
\bibliography{main}

\end{document}